\documentclass[prd,twocolumn,tightenlines]{revtex4}
\usepackage{amsmath}
\usepackage{amssymb}
\usepackage{graphicx}

\newcommand{\be}{\begin{eqnarray}}
\newcommand{\ee}{\end{eqnarray}}

 \newcommand{\gsim}{\mathrel{\hbox{\rlap{\lower.55ex \hbox {$\sim$}}
                   \kern-.3em \raise.4ex \hbox{$>$}}}}
\newcommand{\lsim}{\mathrel{\hbox{\rlap{\lower.55ex \hbox {$\sim$}}
                   \kern-.3em \raise.4ex \hbox{$<$}}}}

\begin{document}

\title{Production of soft $e^+e^-$ Pairs  in Heavy Ion Collisions at RHIC \\ by Semi-coherent Two Photon Processes }
\author{ Pilar Staig  and Edward Shuryak \\ Department of Physics and Astronomy \\ Stony Brook University, Stony Brook NY 11794 USA }
\date{\today}

\begin{abstract}
We calculate the contribution of the two photon production process
into $e^+e^-$ spectra, and compare the results with experimental
data from the PHENIX detector at RHIC. We study the  contribution
given by ``semi-coherent" kinematics, in which one photon is
relatively hard and is incoherently emitted by participating
protons, while another can be soft enough to be in a coherent
domain.

\end{abstract}

\maketitle

\address {Department of Physics and Astronomy, State University of New York,
Stony Brook, NY 11794}

\section{Introduction}

Thirty years ago one of us \cite{Shuryak:1978ij,Shuryak:1980tp}
had suggested to use dileptons and photons as ``penetrating
probes" for dense hadronic matter created in ultrarelativisitc
heavy ion collisions, which -- unlike hadrons -- are observable
from all stages of the collisions and thus can tell us what the initial hottest temperature reached can be.
 It is a very challenging task
for experiments, as one has to remove hadronic backgrounds 
orders of magnitude larger than the photon or dilepton signal. And yet, over
the years there were successful measurements, both at CERN SPS
(muon pairs by NA50/NA60, electron pairs in CERES, photons in WA98) and RHIC (photons,muons and
electrons in PHENIX, electrons in STAR).
We will not go into details of these works, just make few general comments.

 Already  the above mentioned papers from 1970's have singled out the so called
 intermediate mass dileptons (IMD's), with the mass 1-3 $GeV$
 or between $\phi$ and $J/\psi$ resonances, as the window for observing  the thermal QGP radiation.  More detailed
predictions have been made in Ref. \cite{Rapp:1999zw}, where it has also been predicted
that most of those pairs observed are $not$ from
charm decays, as  was widely believed at the time. Only with successful
completion of the NA60 experiment, with its sophisticated charm tracking, this collaboration had
recently  confirmed that they do indeed observe thermal radiation
from QGP \cite{Arnaldi:2008vs} and not just charm decays. For summary of other NA60 results see e.g.\cite{DeFalco:2009em}:
those include dileptons with small masses which come from
resonances $\rho$ mesons decaying  in hadronic and
near-$T_c$ region. Although
still far from being perfect, the existing theory provides a reasonable overall description of the NA60, see
e.g. \cite{Dusling:2007kh,vanHees:2007th}. Important recent observation of   thermal photon radiation from hadronic gas and QGP
has been also made by PHENIX collaboration \cite{:2008fqa}, which is also in overall fair agreement
with the current theory and the hydrodynamical picture of the collision.

\begin{figure}[t]
\includegraphics[width=9 cm]{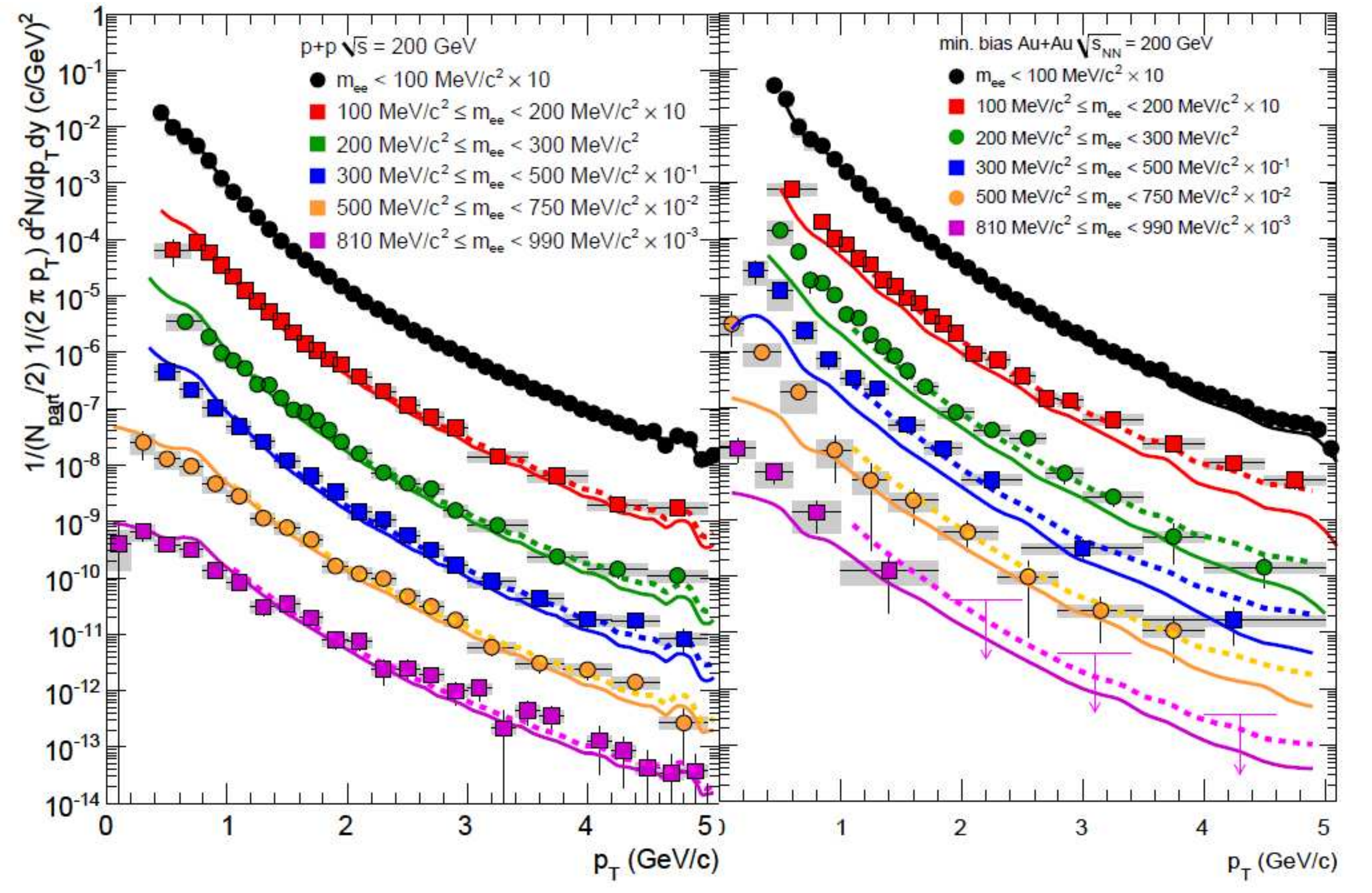}
\vspace{-4ex}\caption{ (Color online) Acceptance corrected invariant $e^+e^-$ yield versus total transverse momentum
of the dilepton pair,  for pp collisions (left) and AuAu collisions (right), from PHENIX publication  \protect\cite{:2009qk}.
The solid curves show the expectation from the sum of the so called
hadronic cocktail contribution plus charm decays. }
 \label{PHENIX}
\end{figure}

And yet, some aspects of the experimental data at RHIC remain
puzzling. Dilepton results from PHENIX show production rate of
small mass $M\sim 500\, MeV$ dileptons few times above theory predictions.
Another puzzle
 is the presence of  the so called ``cold" component in the
dilepton spectrum for  $p_t < 500 \,  MeV$, which is shown in  Fig.\ref{PHENIX}.  While the pp data (points in the l.h.s.)
agree rather well with ``hadronic cocktail" (curves),  in AuAu data (r.h.s.) one finds
 systematic  upward deviations of the data from from similar  curves, at small $p_t$ (the left side of the AuAu plot).
If fitted with exponential, the data  have  a slope
$T_{eff}\approx 100\, MeV$, which is about twice smaller than the
typical slope of the main ``hot" component. What is especially
strange about it is that this slope seems to be the same for
different dilepton mass bins, see the three lowest curves on the
right hand side in. This is in contrast to the ``hot" component,
which shows  $T_{eff}$ increasing with $M$, in good agreement with
expectations based on hydrodynamical picture of expanding matter.
It is a presence of such ``cold" component which originally
motivated us to have a look at some dilepton production mechanisms
which are not included in the ``standard" theory toolbox.

Small mass component is another puzzle, it has unusual centrality
dependence.

Coherent two-photon processes are a  well-known source of small
mass and small $p_t$ dileptons. Their basic theory had been
developed already in 1930's, when Landau and Lifshitz
\cite{Landau:1934zj}  calculated their  total cross section for
nuclear collisions using Weizsacker-Williams (WW) approximation.
There have been extensive studies of the so called ultraperipheral
processes in RHIC environment, for experimental results from STAR
collaboration see \cite{Abelev:2007nb}. As the name suggests,
those processes take place at very large impact parameters $b>2R$,
at which no nuclear interactions take place. For electron pair
production the characteristic $b$ are related to the electron
mass,and are thus very high. Theory development  including all
orders in $Z\alpha$ has been worked out in the last decade.

However the contribution of such processes at near-central
collisions (when multiple hadronic production does happen) and for
the kinematical range of $p_t,M$ seen by PHENIX and NA60 has not
to our knowledge been considered. This is what we are going to do
in this work.

Additional motivation for looking at the two-photon processes
comes from the standard relations between on-shell and slightly
virtual photons $\gamma^*$, which are seen as small-mass
dileptons. PHENIX has used such relations, relating dileptons with
masses $M>100 \, MeV$ with real photons. However, the two-photon
collisions that we discuss do not obey it, producing only
dileptons but not photons, and the question is how important are
those in the kinematical range at hand.
 \section{The formalism}
We use the Equivalent Photon Approximation (EPA)\cite{LL,Jackson}
to determine the differential cross-section for the production of
dileptons in Au-Au collisions. According to this method the
effects of the electromagnetic fields from the moving nuclei can
be replaced by the equivalent photon spectrum
\\
\begin{eqnarray}
dn_i & = & \frac{Z_i^2 \alpha}{\pi^2} \frac{q_{i \perp}^2 \left[F
\left( q_{i \perp}^2 + \frac{w_i^2}{\gamma^2}
\right)\right]^2}{\left(q_{i \perp}^2 +
\frac{w_i^2}{\gamma^2}\right)^2} \frac{d^3q_i}{w_i}
\end{eqnarray}
where $Z_i$ is the number of protons in the nucleus, $F(q^2)$ is
the form factor of the nucleus charge, $q_{i \perp}$ is the
transverse momentum of the photon and $w_i$ is its energy. The
differential cross-section for the gold-gold collision is then
given by the product between the photon spectrum of each nucleus
and the cross-section for the production of dileptons from a
2-photon collision:
\\
\begin{eqnarray}
d \sigma & = & \sigma_{\gamma \gamma} dn_1 dn_2
\\
\end{eqnarray}
which can be written in terms of the total transverse momentum
$\vec{Q}=\vec{q_1}+\vec{q_2}$ and integrated over $\vec{q_2}$ to
give:\\
\begin{eqnarray}
d \sigma & = & \sigma_{\gamma \gamma} \frac{(Z^2 \alpha)^2}{\pi^4}
\frac{dq_{1z} dq_{2z}}{w_1 w_2}d^2 Q_{\perp} \nonumber \\
& & {} \int \frac{q_{2 \perp}^2 \left[F\left(q_{2 \perp}^2 +
\frac{w_2^2}{\gamma^2} \right)\right]^2}{\left(q_{2 \perp}^2 +
\frac{w_2^2}{\gamma^2}\right)^2} \nonumber
\\
& & {} \frac{(\vec{Q_{\perp}}-\vec{q_{2 \perp}})^2
\left[F\left((\vec{Q_{\perp}}-\vec{q_{2
 \perp}})^2 + \frac{w_1^2}{\gamma^2} \right)\right]^2}{\left((\vec{Q_{\perp}}-\vec{q_{2 \perp}})^2 +
\frac{w_1^2}{\gamma^2}\right)^2}d^2 q_{2 \perp}
\end{eqnarray}
\\
Following \cite{Budnev:1974de} we see that the main contributions
to the cross-section come from the regions where $q_1$ and $q_2$
are small (of the order of $w_i/\gamma$). If both of the momenta
are small then the total transverse momentum would also have a
small value and we are interested in studying the dilepton
production for total transverse momentum up to about $0.7$ GeV.
This is why we work in a semi-coherent approach, in which from one
of the nuclei we will get a coherent electric field, which will
correspond to a photon with small transverse momentum, while the
momentum from the other photon can have greater values. This means
that in this case we won't be getting a coherent field from all
the nucleus, but that the protons that compose it can have an
individual effect.  For this case, instead of using the form
factor for a continuous charge distribution we will use the one
coming from considering that the nucleus is composed of Z point
particles.
\\
\begin{figure}[!h]
\begin{center}
\includegraphics[width=8 cm]{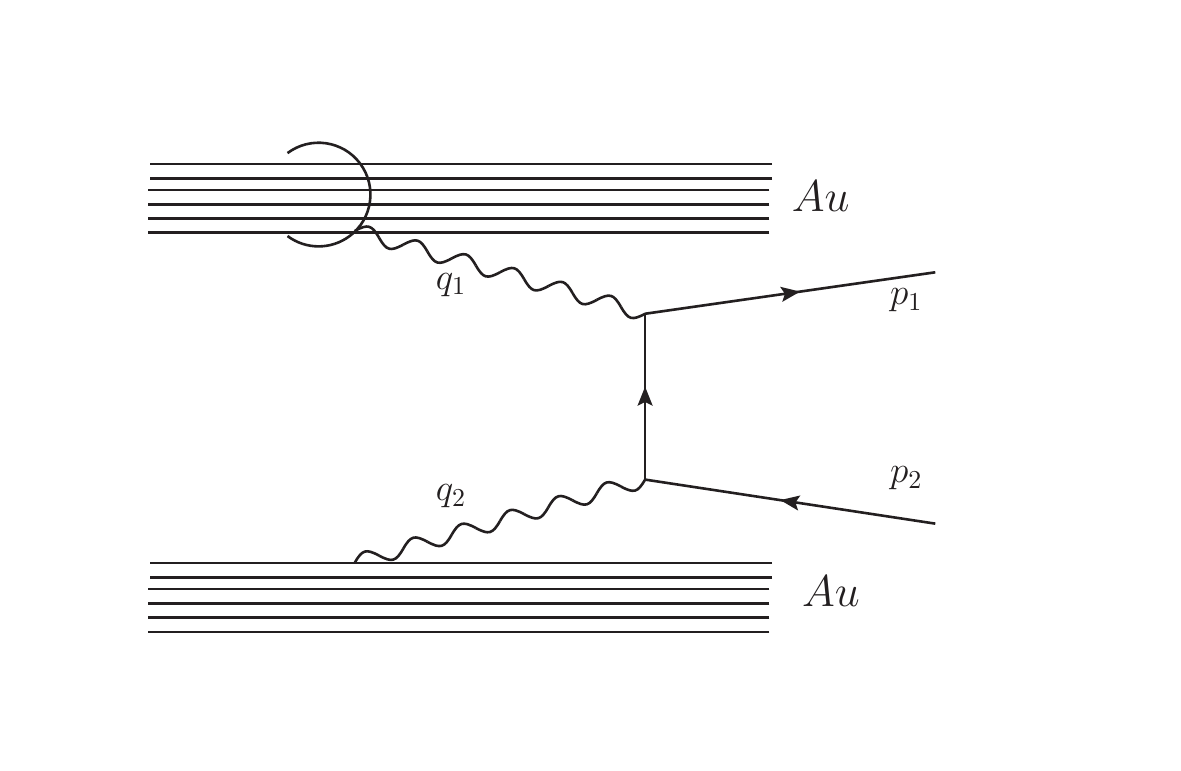}
\end{center}
\vspace{-4ex}\caption{ (Color online)Dilepton production from a semi-coherent
process.}\label{diag}
\end{figure}
\\
For the case when $q_{2\perp}\ll Q_{\perp}$ we can approximate
$\vec{Q_{\perp}}-\vec{q_{2\perp}}\sim \vec{Q_{\perp}}$ so that we
can take all the terms with $Q_{\perp}$ from the integral, to get:
\\
\begin{eqnarray}
d \sigma & = & 2 \sigma_{\gamma \gamma}\frac{(Z^2
\alpha)^2}{\pi^4} \frac{\bm{Q_{\perp}}^2
\left[F\left(\bm{Q_{\perp}}^2 + \frac{w_1^2}{\gamma^2}
\right)\right]^2}{\left(\bm{Q_{\perp}}^2 +
\frac{w_1^2}{\gamma^2}\right)^2}\frac{dq_{1z} dq_{2z}}{w_1 w_2}d^2
Q_{\perp}\nonumber \\
& & {} 2\pi\int \frac{q_{2 \perp}^3 \left[F\left(\bm{q_{2
\perp}}^2 + \frac{w_2^2}{\gamma^2}
\right)\right]^2}{\left(\bm{q_{2 \perp}}^2 +
\frac{w_2^2}{\gamma^2}\right)^2} dq_{2 \perp}
\end{eqnarray}
\\
The factor $2$ in front comes from summing the two cases: when
$q_{1\perp}$ is small and when $q_{2\perp}$ is small.  Now, using
\begin{eqnarray}
w_1+w_2 & = & m_t\cosh{y}\nonumber\\
q_{1z}+q_{2z} & = & m_t\sinh{y},
\end{eqnarray}
where $w_1=\sqrt(Q_{\perp}^2+q_{1z}^2)$, $w_2=|q_{2z}|$and
$m_t=\sqrt{M^2+Q_{\perp}^2}$, we make a change of variables from
the photon longitudinal momenta $q_{1z}$ and $q_{2z}$ to the
invariant mass $M$ and the rapidity $y$. Then, putting $y=0$ we
get:
\\
\small
\begin{eqnarray}
\frac{dq_{1z} dq_{2z}}{w_1 w_2} & = & \frac{M}{\left(
1+\frac{M^2}{\sqrt{4m_t^2Q_{\perp}^2+M^4}}\right)^2}\left(1 +
\frac{\sqrt{4m_t^2Q_{\perp}^2+M^4}}{M^2}\right) \nonumber
\\
& & {} \cdot \frac{4m_t^2}{M^2\sqrt{4m_t^2Q_{\perp}^2+M^4}}dMdy
\nonumber \\
& = & J(M,Q_{\perp})dMdy
\end{eqnarray}
\\ \normalsize
Finally taking the integral over the invariant mass $M$, we get
the cross-section as a function of the total transverse momentum
$Q_\perp$ and the rapidity $y$.
\\ \small
\begin{eqnarray}
\label{yield}\frac{1}{2\pi Q_{\perp}}
\frac{d^2\sigma}{dQ_{\perp}dy} & = & \frac{4 (Z^2\alpha)^2}{\pi^3}
\frac{\left[F(Q_{\perp})\right]^2}{Q_{\perp}^2} \frac{1}{2\pi} \int \int \sigma_{\gamma \gamma}\nonumber \\
& & {}  \int \frac{q_{1\perp}^3\left[F\left(q_{1\perp}^2 +
\frac{w_1^2}{\gamma^2}\right)\right]^2}{\left(q_{1\perp}^2 +
\frac{w_1^2}{\gamma^2}\right)^2} dq_{1 \perp}dM d\phi_Q
\end{eqnarray}
\\ \normalsize
Since we work in the semi-coherent approach the cross-section
$\sigma_{\gamma \gamma}$ is calculated using
$q_1=(w_1,\vec{Q}_{\perp},q_z)$ and $q_2=(q_z,\vec{0},-q_z)$. This
gives as a result $\sigma_{\gamma
\gamma}(M,\vec{Q}_{\perp},\phi_1,\theta_1)$. The angle $\phi_Q$ is
integrated over $2\pi$. The PHENIX detector covers $|\eta|<0.35$
and a total of $180^o$ in azimuth, but the data has been
acceptance corrected to include electrons and positrons from all
directions.  The only restriction that we must impose is due to
the single track acceptance condition that $p_{\perp}>0.2 \,GeV$.
\section{Form Factors}
\begin{figure}[!h]
\begin{center}
\includegraphics[width=8 cm]{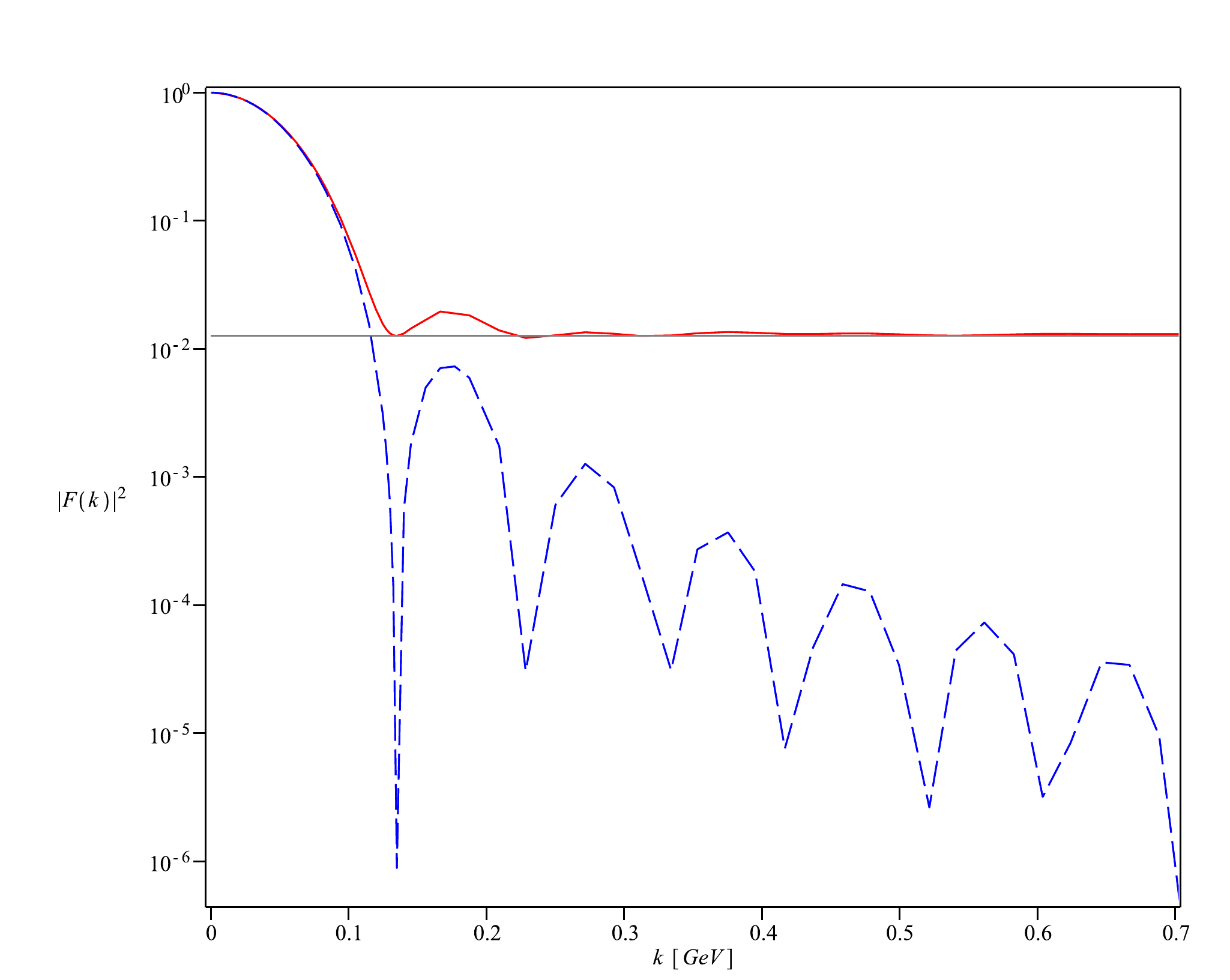}
\end{center}
\vspace{-4ex}\caption{ (Color online)The square of the form factors plotted on a
logarithmic scale. The (blue) dashed line  corresponds to smooth Woods-Saxon charge distribution, the (red) continuous line
corresponds to resolved
discrete protons (but not quarks), as explained in the text.)}\label{FF}
\end{figure}

The charge distribution of the nucleus can be well parameterized
by the Woods-Saxon expression
\\
\begin{eqnarray}
\rho(r) & \propto & \frac{1}{e^{\frac{r-R}{a}}+1}
\end{eqnarray}
\\
with two parameters, the nuclear radius R (6.55 fm for Au)and the
width of the nuclear edge which is typically about $a=0.5$ fm
\cite{nucleardata}. Starting from this charge distribution it is
not possible to get an analytical expression for the form factor,
but the integrals of the fourier transformation can be done
numerically, to get a form factor of the shape seen in FIG.
\ref{FF}.

As it has been stated previously, we are working in a
semi-coherent approach.  This means that while one of the photons
is soft and thus sees the nucleus as a uniform charge distribution
(with the Woods-Saxon shape), the other can have a large
transverse momenta and thus resolve individual protons.  For this
later case we will use the picture of instantaneously frozen
nucleons, which just means that at any given moment the protons
are randomly distributed in the nucleus according to some weight
and frozen in these positions $x_m$, where x is in the direction
where the momentum $Q_{\perp}$ is directed. So the form factor can
be written as:
\\
\begin{eqnarray}
F(k) & = & \frac{1}{Z} \int e^{ikx}\sum_{m=1}^Z \delta(x-x_m)dx  \nonumber
\\
 & = & \frac{1}{Z} \sum_{m=1}^Ze^{ikx_m}
\end{eqnarray}
\\
In the amplitude we have the square of the form factor, so what we
need is:
\\
\begin{eqnarray}
\label{Fdisc}\left|F(k)\right|^2 & = & \frac{1}{Z^2}\left[Z + 2
\sum_{m=1}^Z \sum_{n=1}^{m-1} \cos{\left(k(x_m-x_n)\right)}\right]
\end{eqnarray}
\\
In Fig.\ref{FF} we plot this quantity and compare it to the form
factor from the Woods-Saxon distribution.  We see that for small
momenta they are almost the same but as the momentum is increased
the Woods-Saxon one goes to zero while the one for the point
charges doesn't. This happens because at small momentum the
nucleus is seen as a whole and all the protons are acting
coherently. From eqn.\ref{Fdisc} we see that for $k(x_m-x_n) \ll
1$ the square of the form factor $|F(k)|^2 \sim 1$, which means
that the nucleus acts coherently for momenta up to about $k\sim
1/R$ but as the momentum increases, the contributions from the
different protons start to add up incoherently until, at high
momentum, all the different phases cancel out and we are left only
with protons interacting individually with other protons.  So the
form factor for the Woods-Saxon charge distribution, or for any
other uniform distribution which may represent the nucleus, is
only valid for low momenta but stops working as the momentum
increases, due to the fact that the nuclear substructure  can be
discerned at high enough momenta.

\section{Results}
In order to present our results in a way that allows them to be
compared with data from RHIC, we will study the invariant yield
which is given by:
\\
\begin{eqnarray}
Yield & = & \frac{1}{N_{part}/2}\frac{1}{2\pi Q_{\perp}}
\frac{d^2N}{dQ_{\perp}dy} \nonumber \\
& = & \frac{1}{N_{part}/2}
\frac{1}{\sigma_{total}}\frac{1}{2\pi Q_{\perp}}
\frac{d^2\sigma}{dQ_{\perp}dy}
\end{eqnarray}
\\
\noindent where $\sigma_{total}\sim 4\pi R^2 \sim 1.4\cdot 10^{4}$
[$GeV^{-2}$]. The data from PHENIX correspond to a minimum bias
situation, so the collisions considered have centrality in the
range $0-92\%$ which corresponds to $N_{part}=109$
\cite{Dahms:2008bs}.  Note  that very peripheral, elastic and diffractive events, both in terms of pion production and nuclear physics, are not included:
however those still can generate dileptons.

It is also important to notice that the ``acceptance corrected"
simply means the geometric acceptance of the pair. Separately from
this, the detector has a single track acceptance condition for the
transverse momentum of each lepton \be p_{\perp}>p_{min}=0.2 \,
GeV  \label{condition} \ee needed for them to reach the detector
in the current magnetic field. The reported cross section
\cite{:2009qk} still is under this condition. This  translates
into a truncation for small pair transverse momentum $Q_{\perp}$,
for invariant masses smaller than $0.4 GeV$ .

In our calculation, if the electron mass is neglected, the
$\gamma\gamma\rightarrow e^+e^-$ cross section has collinear
singularity, when the leptons move along the directions of the
photons. The experimental condition (\ref{condition}) supersedes
the cutoff coming from finite electron mass: thus we apply the
same single track acceptance condition  in our calculations.

Since we work in a semi-coherent approach, we keep the transverse
momentum of one of the photons small, with an upper bound given by
$min(Q_{\perp},1/R)$, where the whole nucleus acts coherently.  In
principle, the lower bound on $q_{2\perp}$ is extremely small,
related to the electron mass, and ultraperipheral collisions with
very large impact parameters would be included in the result.
However, as we mentioned earlier, these are not included in the
PHENIX data,  so in order to compare it with our results we had to
restrict the minimum value of $q_{2\perp}$. Fortunately, the
integral over this variable is logarithmic, so we need only
order-of-magnitude estimate of its lowest value. Roughly speaking,
PHENIX only includes collisions in which two nuclei directly touch
each other, or $b < 2R$. The strongest and the  weakest electric
fields of one nuclei inside the disc of the other  corresponds to
distances from $R$ to to $3R$.  The latter distance we thus
associate with the cutoff on the minimal $q_{2\perp}$ value. This
is needed only for the ``softest" photon: the other one has large
$q_{1\perp}$ , by kinematics,  and thus corresponds to field
fluctuations inside the disk of the nucleus 1.

\begin{figure}[!h]
\begin{center}
\includegraphics[width=8 cm]{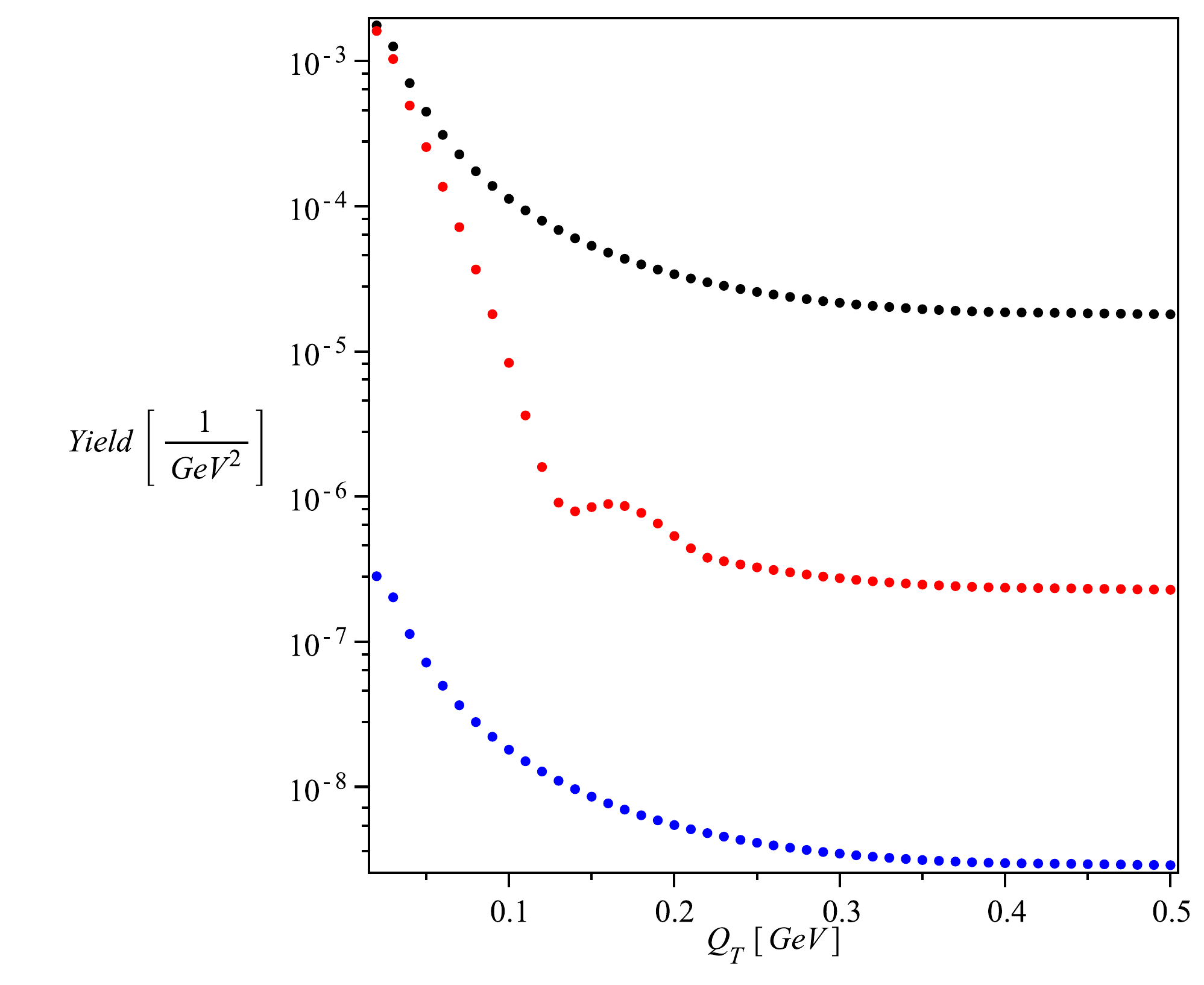}
\end{center}
\vspace{-5ex}\caption{ (Color online)Yield versus total transverse momentum. From
top to bottom: coherent, semi-coherent, totally incoherent.
}\label{co_in}
\end{figure}

In Fig.\ref{co_in} we show our results for the yield as a function
of the total transverse momentum $Q_{\perp}$.  As expected, we see
that for small momenta, while it is in the coherent regime, the
differential cross-section is large, but when the momentum is
increased the yield decreases.  We compare the semi-coherent case
with a totally coherent case, in which we took the nuclei to be
point particles of charge Ze, and with the totally incoherent
case, when we ignore all interferences and consider collisions of
protons each on each. We see that the semi-coherent case lies in
between. It starts, for low transverse momentum, overlapping with
the totally coherent curve and as the momentum increases it drops,
but it never reaches the incoherent curve because we let one of
the photon transverse momentum be small, so that one of the nuclei
is always giving a coherent
contribution.\\
\begin{figure}[!h]
\includegraphics[width=6 cm]{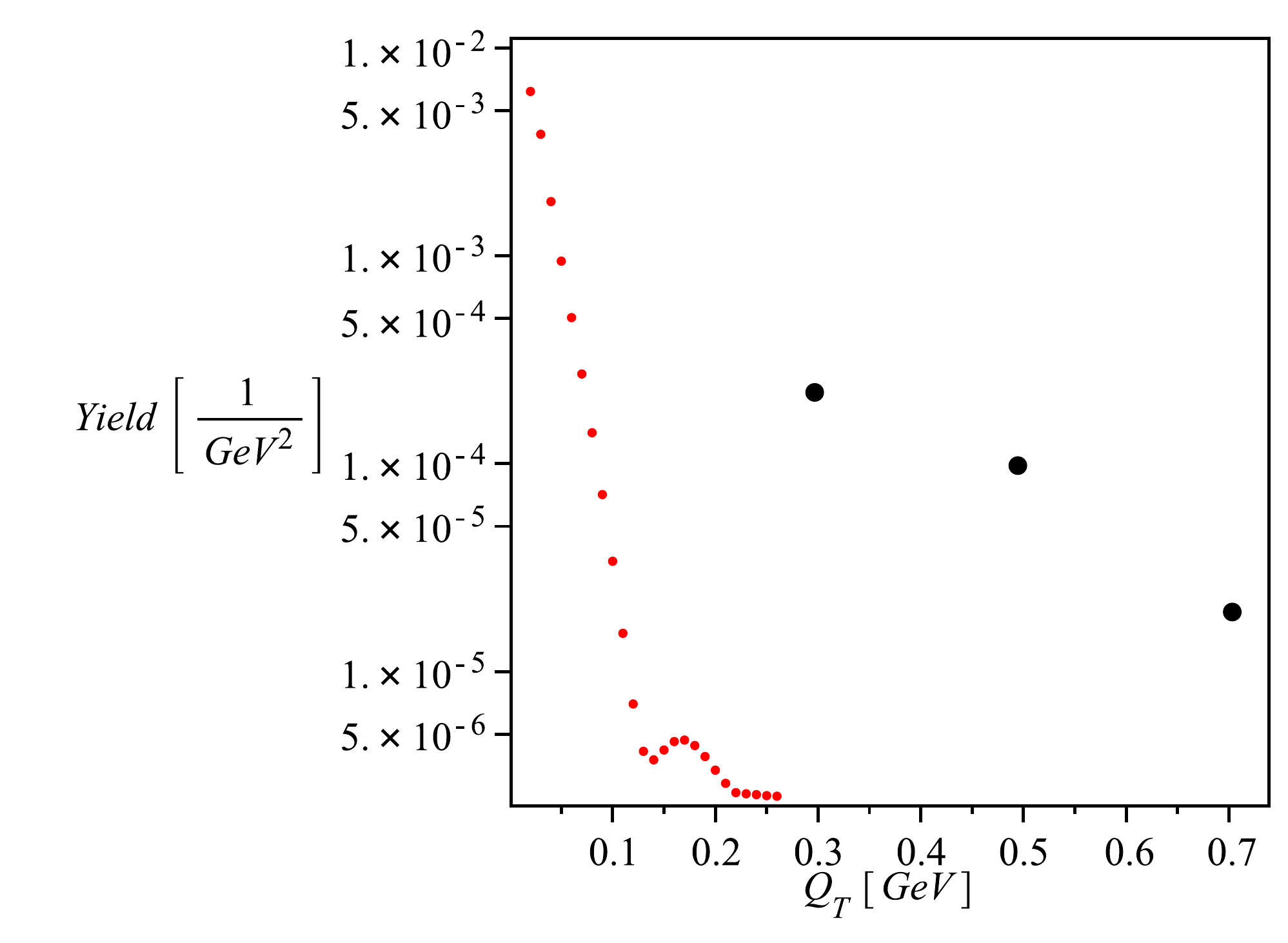}
\includegraphics[width=6 cm]{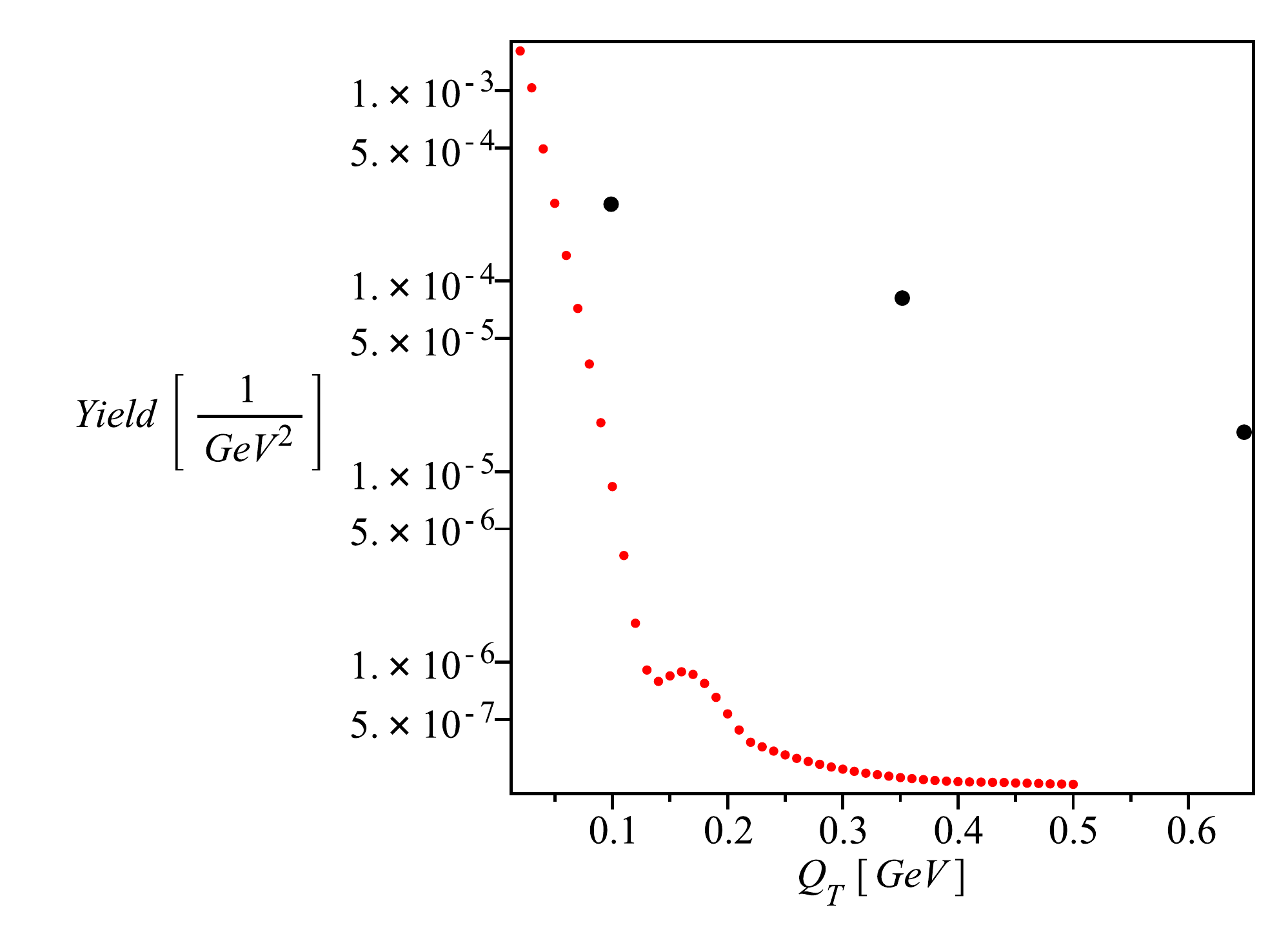}
\includegraphics[width=6 cm]{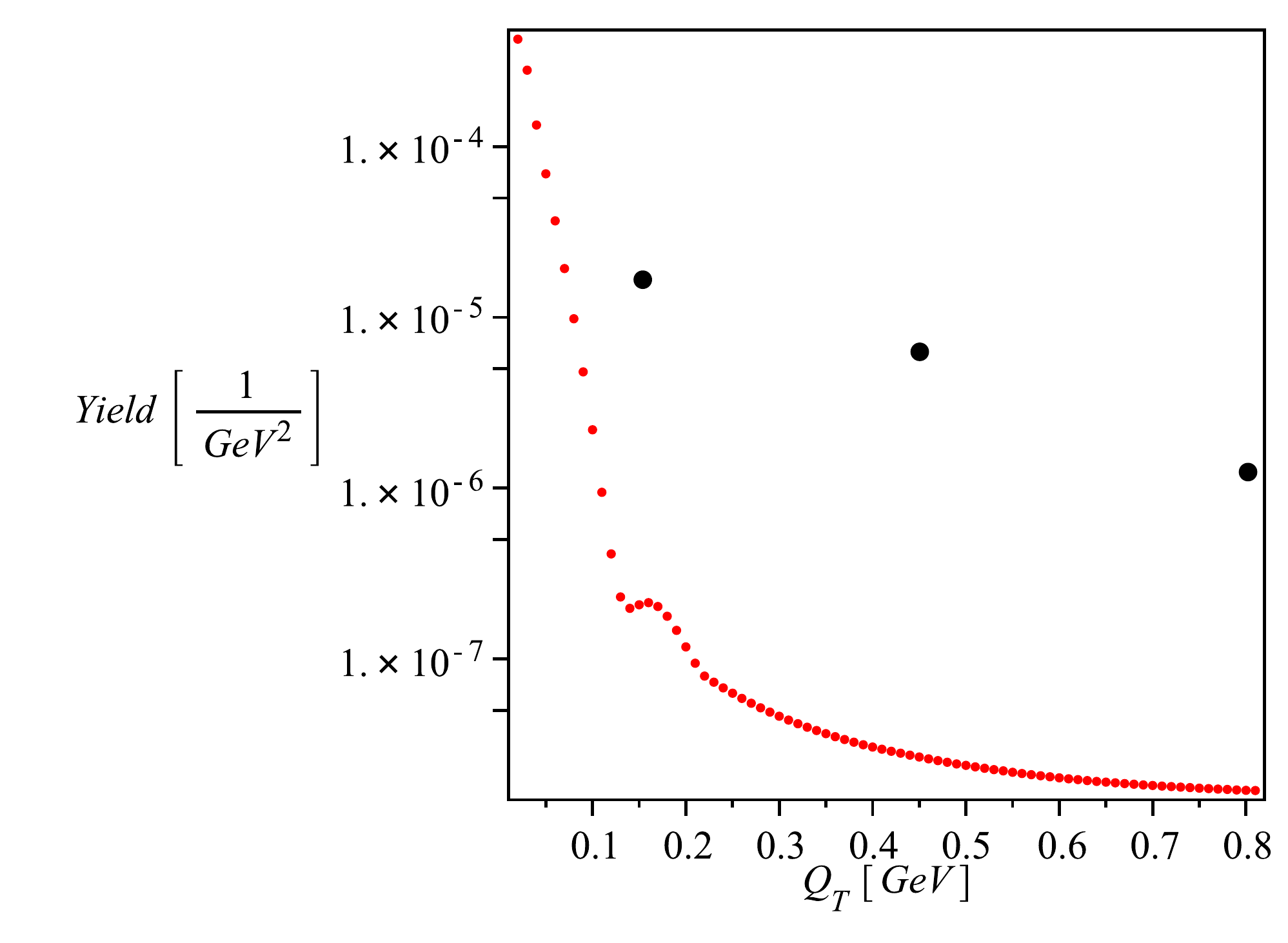}
\vspace{-5ex}\caption{ (Color online)Yield versus total transverse
momentum for different invariant mass ranges. Open (red) symbols
are the two-photon contribution,  compared with PHENIX data (black
closed points, taken from Fig.\ref{PHENIX}). From top to bottom:
M=300-500,500-750, 810-990 MeV. For the lower mass range the
single track acceptance was relaxed from $p_{\perp}>0.2$ to
$p_{\perp}>0.1$ in our calculations.}\label{allmass}
\end{figure}
For different mass bins the shape of the yield as a function of
the transverse momentum is the same.  In Fig.\ref{allmass} the
yield for three mass ranges is presented and it can be seen that
the lowest the invariant mass the higher the yield, as expected.
In order to include the upper plot, which is  in the mass range
$300-500 MeV$, the single track acceptance constraint has been
relaxed to $p_{\perp}>0.1 GeV$ for this case. We compare our
results with experimental data from PHENIX, which is given by the
filled black dots in the plot. We see that the contribution from
the semi-coherent production of dileptons is about two orders of
magnitude smaller than the experimental results. From this we
conclude that in the momentum ranges explored the dilepton
production from two photon collisions doesn't contribute
significantly to the total production of electron-positron pairs
in Au-Au collisions, but this mechanism is important for smaller
transverse momenta, when the two photons are a result of the
coherent interaction of the nuclear electromagnetic fields.

The applicability of the EPA requires that the photon transverse
momenta are small in respect to all other invariants. Therefore,
in  Fig.\ref{allmass} we stop our curves  when $Q_\perp^2$ reaches
the dilepton mass squared, where this condition is not fulfilled.
Note however that in the region where the two-photon production
has chances to be observed, this condition is rather accurate.

For completeness, we also calculated a contribution of this
semi-coherent approach for the LHC energy range, where $\gamma
\sim 3400$. In the unrestricted case, when the photon with small
transverse momentum is in the range
$0<q_{2\perp<min(Q_{\perp},1/R)}$, the integral over $q_{2\perp}$
gives the usual $ln(\frac{\gamma q_{2\perp max}}{w})$, so the
increase in gamma implies a further ``ultraperipheral" enhancement
of the process.  Using the ALICE detector acceptance of
$|\eta|<0.9$ with full range in azimuth and a single track
acceptance of $p_{\perp}>0.1$ \cite{Schicker:2008ih} and allowing
$q_{2\perp}$ to be integrated in the region just described, there
is an enhancement of one order of magnitude in comparison to our
results for the PHENIX acceptance and with $q_{2\perp}$ in the
same range. However, when we restrict $q_{2\perp}$ to be between
$1/3R$ and $min(Q_{\perp},1/R)$ and in this way don't consider
ultraperipheral collisions, the results for the yields that we
calculate for PHENIX and ALICE are very similar and the small
difference between them (about a factor $2$ for small transverse
momentum) is due to the greater acceptance of the latter and also
to the different elements used (Z=79 for RHIC and Z=82 for LHC).
\section{Resolving Quarks}
To end this study of dilepton production in heavy ion collisions,
we consider the structure of the nucleons, this is we resolve
partons such that, as before, from one of the nuclei we get a
coherent contribution (small momenta) while from the other one we
get the effect from partons acting individually. To determine the
parton contribution we must turn to use the parton distribution
functions (PDF's).

The nucleons are composed of quarks and gluons and the probability
that a given nucleon contains a constituent particle with $x$
momentum fraction of the total momentum of the nucleon corresponds
to $f_i(x)dx$, where the functions $f_i$ are the PDF's for the i
type constituent
($i=u,\bar{u},d,\bar{d},s,\bar{s},c,\bar{c},b,\bar{b},t,\bar{t},gluon$)
for a proton.

In our calculations we use data from the CTEQ collaboration
\cite{Pumplin:2002vw} and we only consider the three lightest
quarks: u and d valence quarks and u,d and s sea quarks. To
determine the number of each kind of quark present in a nucleon we
integrate the corresponding PDF's from $x_{min}$ to 1, some
examples can be seen in table \ref{table}.
\begin{table}
\begin{tabular}{|c|c|c|c|c|c|c|c|}
  \hline
   $x_{min}$ & u & $\bar{u}$ & $u_v$ & d & $\bar{d}$ & $d_v$ & s \\
  \hline
   0.01  & 2.452 & 0.763 & 1.690 & 1.703 & 0.864 & 0.839 & 0.521\\
   0.0025 & 4.305 & 2.386 & 1.918 & 3.475 & 2.505 & 0.970 & 1.775\\
   0.001 & 5.249 & 3.253 & 1.996 & 4.390 & 3.377 & 1.013 & 2.451\\
  \hline
\end{tabular}
\vspace{-2ex}\caption{ (Color online)Number of quarks calculated by integrating
the PDF's from CTEQ from different $x_{min}$.}\label{table}
\end{table}

In order to consider all the partons that are capable of emitting
a photon of energy $w$, we take $x_{min}$ to correspond to that
energy. Since in Au-Au collisions at RHIC the typical center of
mass energy per nucleon is of 100 GeV, $x_{min}$=0.01 implies that
all the partons that can emit a photon of $w=1GeV$ are taken into
account. We consider partons from protons and neutrons, the PDF's
for both types of nucleon are related by isospin symmetry so for
the neutrons the up and down quarks are interchanged with respect
to protons.

\begin{figure}[!h]
\includegraphics[width=4.1 cm]{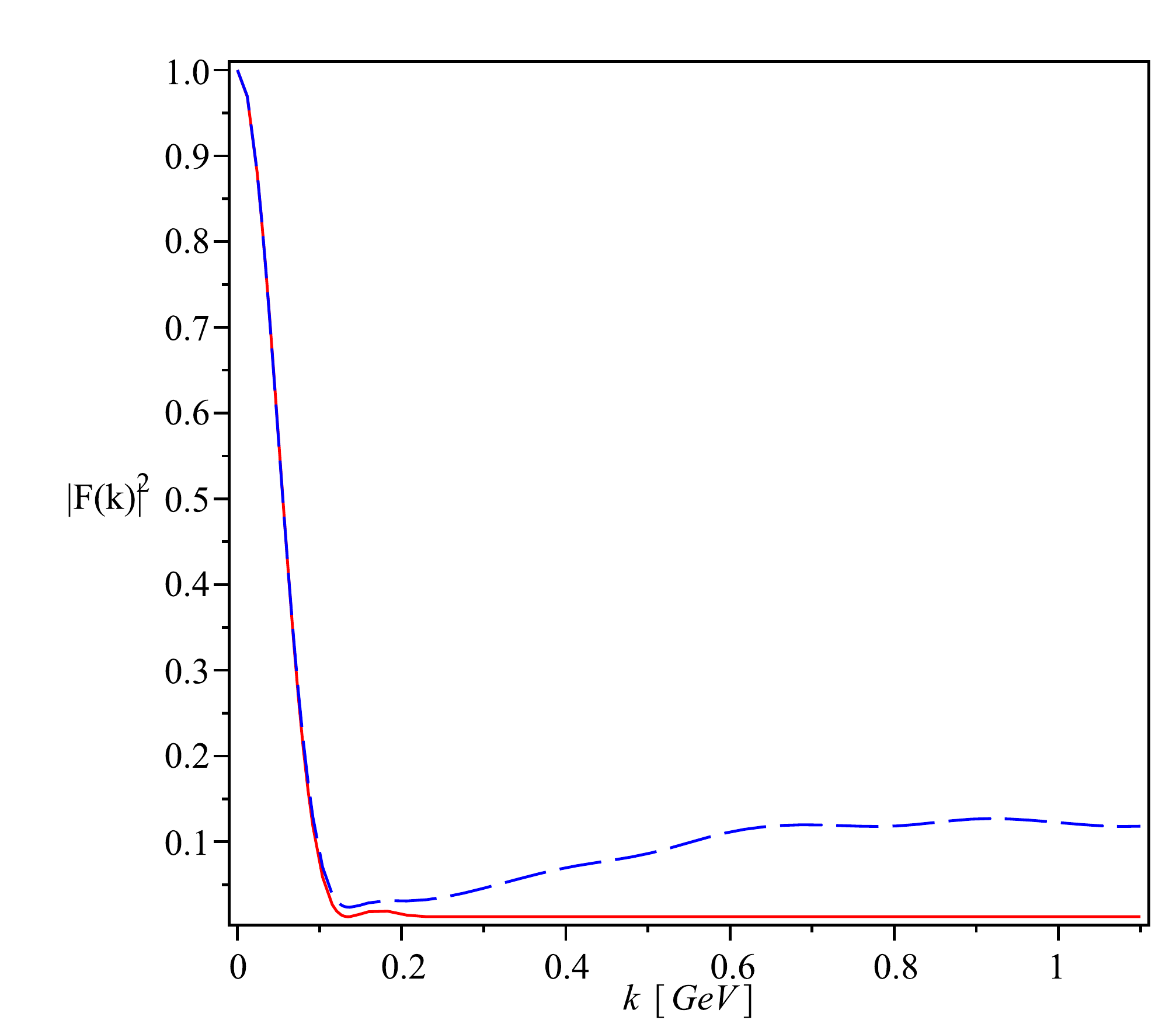}
\includegraphics[width=4.3 cm]{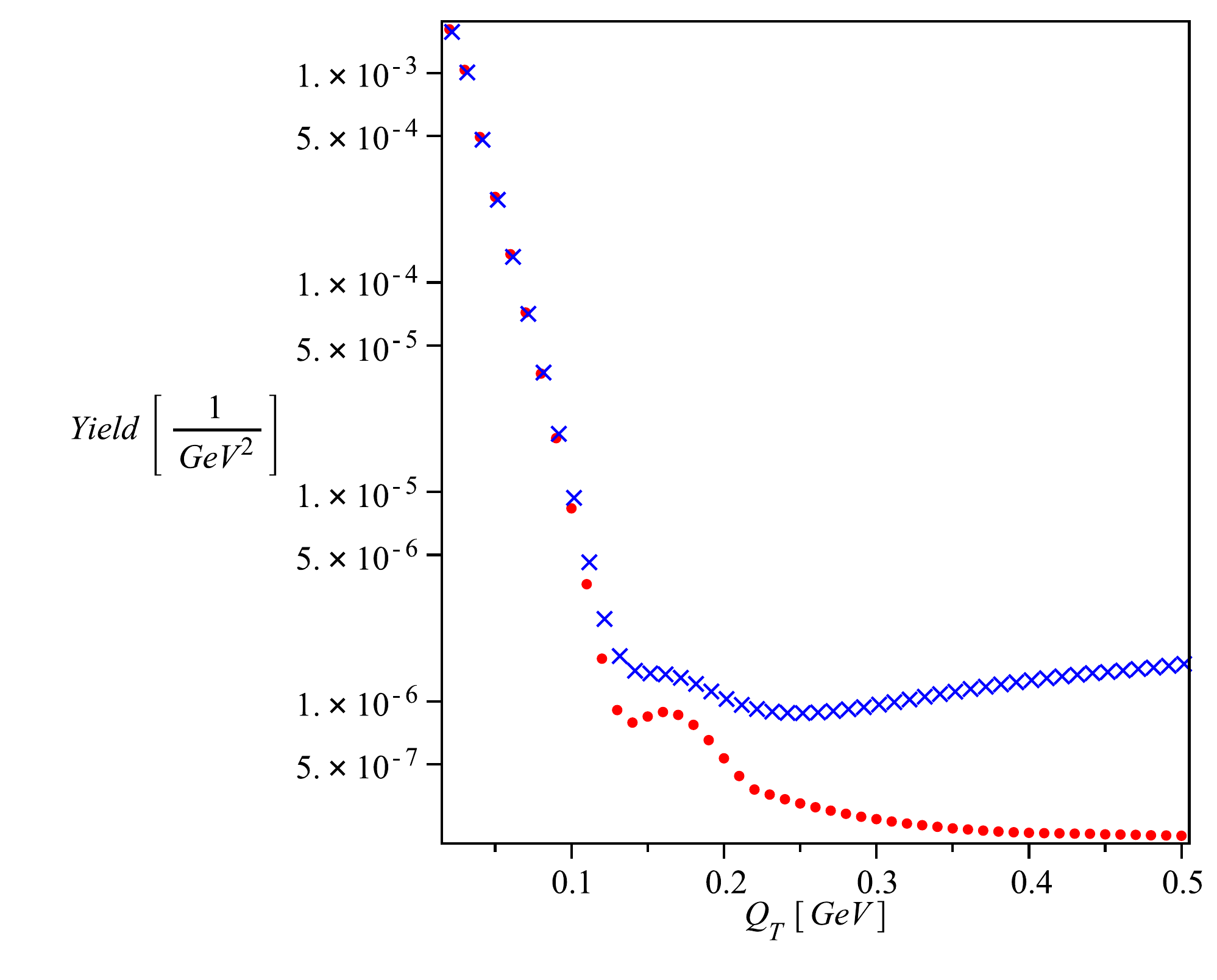}

\vspace{-3ex}\caption{ (Color online)LEFT:Comparison between form factors.
Continuous line:resolving up to protons, dashed line:resolving up
to partons. RIGHT:Yield versus total transverse momentum for an
invariant mas range of 500-750 MeV. Upper curve: resolving up to
partons. Lower curve: resolving up to protons. }\label{quarkFF}
\end{figure}

We use $x_{min}=0.0025$ to  calculate the number of each kind of
quark present in a gold nucleus and then we proceed as before,
working in the instantaneously frozen picture.  We randomly select
the positions of the quarks in the nucleus to determine
\footnotesize
\begin{eqnarray}
F(k)  = \frac{\frac{2}{3}\displaystyle\sum_{i=1}^{n_u}
e^{ikx_{u_{i}}} -\frac{2}{3}\displaystyle\sum_{i=1}^{n_{\bar{u}}}
e^{ikx_{\bar{u}_{i}}} -
\frac{1}{3}\displaystyle\sum_{i=1}^{n_{d}+n_{s}} e^{ikx_{d_{i}}} +
\frac{1}{3}\displaystyle\sum_{i=1}^{n_{\bar{d}}+n_{\bar{s}}}
e^{ikx_{\bar{d}_{i}}}}{\frac{2}{3}(n_u - n_{\bar{u}}) -
\frac{1}{3}(n_d - n_{\bar{d}})-\frac{1}{3}(n_s - n_{\bar{s}})} \nonumber
\end{eqnarray}
\normalsize

\noindent then this quantity is squared and averaged over
different quark configurations.  The result is plotted on the left
side in Fig.\ref{quarkFF} and compared to the previous case when
we resolved only up to nucleons.
For small momenta the form factor is as before, and the nucleus
gives a coherent contribution, but for large momenta, when the
value  comes  from the incoherent contributions of individual
particles, the result obtained from resolving up to partons is
about one order of magnitude  larger than when resolving only up
to protons, because the number of participating particles is
increased since we not only consider the valence quarks from the
proton but also the sea quarks and  all the quarks from the
neutrons.

In Fig.\ref{quarkFF} we also present the comparison for the yield
as a function of total transverse momentum between the two cases
that we have studied: resolving up to protons and resolving up to
partons.  We see that for momenta larger than 0.1 GeV, the
contribution to the dilepton production process is larger when we
take into account the nucleon structure, but it is still orders of
magnitude below the experimental data.

\section{Summary}
We studied the production of dileptons in heavy ion collisions
using the two-photon mechanism, in the double effective photon approximation.
It  is well known since 1930's \cite{Landau:1934zj} that for low
momenta the contribution from both nuclei is coherent and in this
case the cross section for the process is proportional to
$(Z\alpha)^4$.  Our main interest was to look at the regime in which this coherence is lost.

Although our original motivation was to explain the RHIC puzzles, we learned (relatively early in
the calculation) that the two-photon mechanism unfortunately cannot explain any of them. Yet we persisted and completed this calculation,
for two reasons.

One is that with relatively minor modification of the PHENIX
experimental condition -- in particular with reduction of the
magnetic field -- the dilepton from the two-photon mechanism would
become detectable and maybe even dominant. Thus it would be
beneficial to identify  and study those, in next RHIC runs. It
would not require significant expense, as we speak about quite
large cross sections.

Another reason is a theoretical curiosity: what exactly happens
with the two-photon cross section when the coherence is lost. We
found that when it happens, the cross section is dominated by a
``semi-coherent" regime, in which the momentum of one of the
photons remains small  enough to be represented as a coherent
field of one of the nuclei, while allowing the other photon to
have larger momentum and resolve individual particles, protons or
even quarks. We have shown in this paper that the semi-coherent
approach gives  a greater contribution than the totally incoherent
case, while still allowing the dilepton pair to have relatively
large total transverse momentum.

We studied two cases, first resolving only up to protons an then
resolving them into charged partons (quarks).  By comparing both
results we see that the larger the number of particles that are
resolved, the greater the contribution.  Unfortunately, such
increase-- by almost one order of magnitude --  is not enough to
explain the total dilepton spectrum observed at RHIC.

However, there can be similar semi-coherent regime for other
processes in QCD, especially at $Q\sim 1 GeV$.  Traditionally, few-GeV momenta transfer is
seen as a domain of parton model, with its treated via completely incoherent PDF's
and cross sections for hard processes.  And yet, even for gluons, people introduced
``color glass" fields collectively generated by many hard partons \cite{McLerran:1993ni}.
%
The so called ``ridge"
phenomenon, recently discovered by STAR collaboration \cite{star_ridge} , is presumably due to
early local large-scale fluctuation of color field \cite{Dumitru:2008wn}. A collision of such fields with the
usual incoherent partons can presumably generate a QCD analog of ``semi-coherent" processes we studied above.

\section{Acknowledgements}
  We thank Tony Baltz for a useful discussion of the form factors. The
 work is partially
supported by the US-DOE grants DE-FG02-88ER40388 and
DE-FG03-97ER4014.


\end{document}